\documentclass[12pt]{article} 
\usepackage{graphicx} 
\usepackage{amsmath,caption,threeparttable}  
\usepackage{newtxtext,newtxmath} 
\usepackage[T1]{fontenc} 
\usepackage{ae,aecompl} 

\begin{document}
	
\title{{\bf Simple model of violent relaxation \\ 
		of a spherical stellar system }} 
\author{V.~Yu.~Terebizh\thanks{E-mail: valery@terebizh.ru} \\ 
	\small{Crimean Astrophysical Observatory}} 
	
\maketitle 
	
\begin{abstract} 
	
A spherical system with arbitrary mean velocities in the radial and 
transverse directions, but with zero dispersion of radial velocities, 
is considered as a model describing the violent relaxation of star 
clusters and galaxies. The model allows to reduce the infinite 
chain of Jeans' equations of moments to just three hydrodynamic 
equations. The found system of equations is consistent with the law 
of conservation of energy and the virial equation. This part of the 
work is devoted to a general description of the model, then the 
calculation results will be presented. 
\end{abstract} 
	

\section{Introduction} 

The dynamic evolution of stellar systems is usually divided into 
at least two stages: an initial period of ``violent relaxation'' 
(Lynden-Bell 1967), followed by a much longer transition of the 
system to a quasi-stationary state (Spitzer 1987). The latter is not 
a state of thermodynamic equilibrium, which is unattainable both due 
to the openness of astronomical systems and due to the specificity of 
gravitational interaction. It is worth noting that the modeling of 
these periods of evolution is essentially unequal: there are a number 
of analytical and numerical models of quasi-equilibrium evolution, 
whereas the first period was studied almost exclusively by direct 
numerical calculations of the motion of individual stars. The reason 
is that even for the relatively simple geometry of spherically 
symmetric systems, the kinetic equation for the density 
$f(r,v_r,v_t,t)$ in the 3-dimensional phase space formed by the 
radius~$r$, the radial velocity $v_r$, and the transverse 
velocity $v_t$, 
$$
\frac{\partial f}{\partial t} + v_r\,\frac{\partial f}{\partial r} 
+ \left( \frac{v_t^2}{r} - \frac{\partial \Phi}{\partial r} \right) 
\frac{\partial f}{\partial v_r} - \frac{v_r v_t}{r}\, 
\frac{\partial f}{\partial v_t} = 0, 
\eqno(1)
$$ 
accompanied by Poisson equation for the smoothed gravitational 
potential $\Phi(r,t)$, 
 
$$
\frac{1}{r^2}\,\frac{\partial}{\partial r} \left( 
r^2\,\frac{\partial \Phi}{\partial r}  \right) = 
4\pi G m\,n(r,t), \qquad n(r,t) = \int f d{\bf v}, 
\eqno(2) 
$$
remains too complex for analytical study. Eq.~(1) is most often 
referred to as the ``collisionless Boltzmann equation'' (Jeans 1915, 
1919; Shiveshwarkar 1936; H\'enon 1982). 

For simplicity, we assumed above that all stars have the same 
mass~$m$, so the phase density is normalized to the total number of 
stars in the system $N$. Then the Poisson equation can be written as 
$$ 
\dfrac{\partial \Phi}{\partial r} = G\, \dfrac{M(r,t)}{r^2}, 
\quad \mbox{where} \quad  
M(r,t) = 4\pi m \int_0^r  n(r',t)r'^2 dr' 
\eqno(3) 
$$ 
is the instantaneous mass inside a sphere of radius~$r$. 

Jeans' (1915) approach to the problem involves using not the phase 
density, but only its moments with respect to velocities (see, e.g., 
Chandrasekhar 1942; Binney~\& Tremaine 2008). For the case of a 
spherically symmetric system considered here, the moments of 
orders $j,k = 0,1,2,\ldots$ are defined as follows: 
$$  
s_{jk}(r,t) \equiv \int v_r^j v_t^k f d{\bf v} =  2\pi 
\int_{-\infty}^{\infty} v_r^j dv_r \int_0^\infty v_t^{k+1} f dv_t. 
\eqno(4)
$$ 
The special notations are used frequently for some moments, namely, 
the spatial density $n(r,t) = s_{00}$, the flux $u(r,t) = s_{01}$, 
the moments of second order $p(r,t) = s_{20}$ and $q(r,t) = s_{02}$. 
For further, we need only the first~5 equations from the infinite 
chain of Jeans equations of moments that follow from Eqs.~(1) 
and~(4): 
$$ 
\left \{    
\begin{array}{lllll} 
    \dfrac{\partial n}{\partial t} + \dfrac{1}{r^2} 
    \dfrac{\partial}{\partial r} (r^2 u) = 0, \vspace{2mm} & \\ 
\dfrac{\partial u}{\partial t} + \dfrac{1}{r^2} 
\dfrac{\partial}{\partial r} (r^2 p) - \dfrac{q}{r} 
= -n\, \dfrac{\partial \Phi}{\partial r}, \vspace{2mm} & \\ 
    \dfrac{\partial s_{01}}{\partial t} + 
    \dfrac{1}{r^2} \dfrac{\partial}{\partial r} (r^2 s_{11}) + 
    \dfrac{s_{11}}{r} = 0, \vspace{2mm} &  \\  
\dfrac{\partial p}{\partial t} + \dfrac{1}{r^2} 
\dfrac{\partial}{\partial r} (r^2 s_{30}) - \dfrac{2 s_{12}}{r}   
= -2u\, \dfrac{\partial \Phi}{\partial r}, \vspace{2mm} 	& \\ 
    \dfrac{\partial q}{\partial t} + 
\dfrac{1}{r^2} \dfrac{\partial}{\partial r} (r^2 s_{12}) + 
\dfrac{2 s_{12}}{r} = 0.   \vspace{2mm} \\ 
\end{array}    
\right. 
\eqno(5) 
$$
Usually the equations written here are applied individually, whereas 
the whole chain of equations is not closed in the sense that each 
equation of a given order includes terms of a higher order. 

The system of Jeans equations can be closed by choosing some 
special model of the stellar system. The simplest example is the 
classic McCrea~\& Milne~(1934) model, which imitates homogeneous 
cosmological model in Einstein general theory of relativity. 
Considering a homogeneous sphere of radius $a(t)$ and fixed mass~$M$ 
expanding radially according to the Hubble law, the authors arrived 
at the fundamental Friedmann equation 
$$
\left( \dfrac{da}{dt} \right)^2 = \,
\dfrac{8\pi G}{3}\, \rho a^2 + \mbox{const}  \vspace{2mm}, 
\eqno(6)
$$
describing the expansion of the Universe (see Weinberg 2008, 
Eq.~(1.5.19)). In the context of the approach considered here, the 
assumptions of McCrea and Milne imply the following form of the 
phase density: 
$$ 
  \left\{ \begin{array}{ll} 
f(r,v_r,v_t,t) = n(r,t)\cdot\delta[v_r - H(t)\,r]\, 
\delta(v_\theta)\,\delta(v_\varphi)\,,
 \vspace{2mm}&  \\ 
n(r,t) = 3 M/\left( 4\pi m a^3(t) \right), 
\quad 0 \le r \le a(t), 
\vspace{2mm} &\\ 
n(r,t) = 0, \quad r > a(t),  
\end{array}  \right. 
\eqno(7) 
$$
where $\delta(.)$ is the Dirac $\delta$-function, $H(t)$ is the 
Hubble parameter, $v_\theta$ and $v_\varphi$ are the transverse 
velocities in the corresponding directions. Successive substituting 
these expressions into Eqs.~(4) and~(5) leads to the above Friedmann 
equation. 

Of course, the McCrea~-- Milne model is too simple to describe the 
violent relaxation of spherical stellar systems such as globular 
clusters and galaxies, where the density distribution is significantly 
non-uniform and the expected velocity distribution should differ 
radically from the Hubble law. At the same time, it can be hoped that 
a model with an arbitrary density distribution and a moderate 
simplification of the nature of the movements will make it possible 
to reveal some essential details of real dynamic evolution. This is 
precisely the attempt made in this note: we assume zero velocity 
dispersion in the radial direction and some simple form of the 
distribution of transverse velocities, leaving the mean velocities 
arbitrary. In this case, the open system of Eqs~(5) is reduced to a 
closed system of only three hydrodynamic equations.

\section{Model with non-dispersive radial motion} 

The proposed model is based on the following representation of 
the phase density: 
$$ 
\begin{array}{ll} 
	f(r,v_r,v_t,t)d{\bf r}d{\bf v} = \\ 
	n(r,t)\, 4\pi r^2 dr \cdot \delta[v_r - v(r,t)]dv_r \cdot 
	\exp{(-v_t^2/2s^2)}\, \dfrac{v_t dv_t}{s^2(x,t)}\,. 
\end{array}  
\eqno(8) 
$$
As can be seen, the mean radial velocity is described by some 
function $v(r,t)$, whereas in relation to the transverse velocities 
only the distribution shape is specified, namely, a Gaussian 
distribution with zero mean and standard deviation $s(r,t)$ along 
each of the two orthogonal directions $\theta$ and $\varphi$. 
Since $v_t^2 = v_{\theta}^2 + v_{\varphi}^2$, the distribution of 
transverse velocity $v_t$ is then given by the manner specified 
above (see Appendix). Thus, the phase density of the system is 
given by three functions: $n(r,t)$, $v(r,t)$, and $s(r,t)$. 

Further consideration involves finding from Eq.~(4) the moments 
corresponding to the density (8), and then substituting the found 
moments into Eq.~(5). These technical transformations are given 
in the Appendix. As expected, the continuity equation 
$$
\dfrac{\partial n}{\partial t} + \dfrac{1}{r^2} 
\dfrac{\partial}{\partial r} (r^2 n\, v) = 0 
\eqno(9)
$$ 
is preserved, while the remaining four relations written out 
explicitly in Eq.~(5) are reduced to  
$$
\dfrac{\partial v}{\partial t} + 
v\,\dfrac{\partial v}{\partial r} - 
\dfrac{2 s^2}{r} = -\,\dfrac{\partial \Phi}{\partial r} 
\eqno(10)
$$
and 
$$
\dfrac{\partial s}{\partial t} + 
v\,\dfrac{\partial s}{\partial r} + \dfrac{vs}{r} = 0. 
\eqno(11) 
$$
In view of Eq.~(3), the last three equations form a closed system 
with respect to the sought functions $n,v,s$.  

Before moving on, let us bring the relations we need to a form 
more convenient for numerical solution. The point is that 
Eqs.~(9)~-- (11) are partial differential equations (PDE) in time, 
whereas Poisson's Eq.~(3) has an integral form. The necessary 
transformations are obvious in the case of spherical symmetry. 
Indeed, in view of Eq.~(9) we have:
$$
  \partial M(r,t)/\partial t = -\,4\pi m r^2 n(r,t) v(r,t), 
  \vspace{2mm}
  \eqno(12)
$$ 
which in combination with Eq.~(3) gives us the PDE for the 
instantaneous mass $M(r,t)$. Together with Eqs.~(10) and 
(11) we obtain a closed system of equations for finding 
functions $M(r,t)$, $v(r,t)$, and $s(r,t)$: 

$$  
\left \{ 
\begin{array}{lll} 
  \dfrac{\partial M}{\partial t} + 
v\,\dfrac{\partial M}{\partial r} = 0, \vspace{2mm} & \\ 
  \dfrac{\partial v}{\partial t} + 
v\,\dfrac{\partial v}{\partial r} - 
\dfrac{2 s^2}{r} = -\,\dfrac{GM}{r^2}, \vspace{2mm} & \\ 
  \dfrac{\partial s}{\partial t} + 
v\,\dfrac{\partial s}{\partial r} + \dfrac{vs}{r} = 0.  \vspace{2mm} & \\ 
\end{array} 
\right. 
\eqno(13) 
$$
Of course, the appropriate initial data and boundary conditions 
must be added. It is desirable that the initial configuration 
does not contain sharp changes in density and speed. 

Note that the continuity equation in the form Eq.~($13_1$) is 
preferable in calculations compared to Eq.~(9), since the mass 
$M(r,t)$ does not behave as chaotically in time as the local density 
$n(r,t)$. As for the Eqs.~($13_2$) and Eq.~($13_3$), the nonlinear 
quadratic second terms attract attention because they are 
characteristic of continuum mechanics. Their nature is clearly 
explained, in particular, by Feynman lectures on physics (Feynman, 
Leighton~\& Sands 1964, Ch.~40.2). In general, the last two 
equations of system (13) represent a specification of the classical 
Euler equations (Landau~\& Lifshitz 1987, Ch.~1) to the conditions 
of a stellar system. 

In the model considered here, the kinetic $K$ and potential $W$ 
energies are given as follows: 
$$ 
\begin{array}{ll}
  K(t) =  2\pi m \int_0^\infty [v^2(r,t) + 2s^2(r,t)]n(r,t) r^2 dr,   
      \vspace{2mm} & \\
  W(t) = -\,4\pi m G  \int_0^\infty M(r,t) n(r,t) r dr\,.
\end{array} 
\eqno(14) 
$$ 
It is easy to verify that Eqs.~(13) are consistent both with the 
law of conservation of energy of the system, 
$E = K(t) + W(t) = \mbox{const}$, and with the virial equation 
$$
\dfrac{1}{2}\,\dfrac{d^2 J(t)}{dt^2} = 2E - W(t),  
\eqno(15)
$$
where 
$$
  J(t) \equiv 4\pi m \int_0^\infty n(r,t) r^4 dr 
  \eqno(16) 
$$ 
is the moment of inertia of the system.

\section{Conservation of angular momentum} 

As is known, when a body moves in a central field, its angular 
momentum is preserved (see, e.g., Landau~\& Lifshitz 1976, Ch.~1.9). 
In the case under consideration, this follows directly from Eq.~(1) if 
we write out its characteristic equations; the corresponding integral 
of motion is 
$$ 
  \mu \equiv r v_t = \mbox{const}. 
  \eqno(17)
$$
Further, the presence of the integral of motion allows to simplify 
the original differential equation, so that Eq.~(1) takes 
the form:
$$
\frac{\partial f_\mu}{\partial t} + v_r\,\frac{\partial f_\mu}
{\partial r} + \left( \frac{\mu^2}{r^3} - \frac{\partial \Phi_\mu} 
{\partial r} \right) \frac{\partial f_\mu}{\partial v_r} = 0, 
\eqno(18)
$$ 
where $f_\mu(r,v_r,t)$ and $\Phi_\mu(r,t)$ are the phase density 
and the smoothed potential at a fixed value of angular momentum. 
The potential is related to the spatial density $n_\mu(r,t)$ by the 
Poisson equation, similar to Eq.~(2). 

We recall these facts because in relation to Eq.~(18) one can 
perform operations similar to those made in relation to Eq.~(1). 
Obviously, the Jeans moment system will now be much simpler than 
given by Eq.~(5). It would seem that equations similar to Eqs.~(13) 
would acquire a more attractive form for calculations. However, it 
should be borne in mind that Eq.~(18) describes the dynamics of the 
system {\it not in the entire phase space, but only in hypersurface 
specified by Eq.~(17)}. Such a description is of theoretical 
interest, but in real stellar systems there is a wide range of 
values of angular momentum, while superposition of solutions with 
different moments is impossible due to the nonlinearity of the 
kinetic equation.

\section{Concluding remarks} 

Modern means of numerical solution of partial differential equations 
make it possible to investigate solutions of Eqs.~(13) for a set of 
initial configurations, thereby revealing the characteristic features 
of the violent relaxation of stellar systems. Of primary interest is 
the estimation of the time of the violent relaxation stage $T_v$, 
when the system, after a series of large-scale oscillations, becomes 
sufficiently close to its quasi-stationary state, in other words, 
the virial ratio $V(t) \equiv 2K(t)/|W(t)|$ has become close to~$1$. 
According to von~Hoerner~(1957) and Lynden-Bell~(1967), at the end 
of this stage the spatial density distribution can be expected to 
acquire the characteristic two-component form, while the velocity 
distribution to become close to Maxwellian. The conclusion of 
Terebizh~(2024) that the root-mean-square system radius, 
$R(t) \equiv [J(t)/M]^{1/2}$, and its harmonic radius, 
$\mathfrak{R(t)} \equiv GM^2/2|W(t)|$, behave essentially 
differently also requires verification. 

It should be added that the approach described above represents 
a fairly simple version of the transition from the Jeans infinite 
moment chain to the closed system of hydrodynamic equations. 
The introduction of a non-zero radial velocity dispersion and a 
connection between the radial and transverse velocity components 
into Eq.~(8) seems promising first of all. Much depends on the 
capabilities of the phase density approximation presented here.

%

\section*{Data availability} 

No new data were generated or analysed in support of this research. 

\newpage 

\section*{Appendix. Derivation of hydrodynamic equations} 

First of all, let us present the necessary facts from probability 
theory (Feller 1966, v~2, Ch.~2). 

Let $V_\theta$ and $V_\varphi$ be normal random variables with 
zero means and equal standard deviations $s$. In this case, the 
square $V_t^2 \equiv V_\theta^2 + V_\varphi^2$ is a random variable 
with probability density $\alpha \exp(- \alpha x)$, where 
$\alpha = 1/2s^2$. The random variable $V_t$ itself is distributed 
with density $2\alpha x \exp(- \alpha x^2)$. The corresponding 
averaging gives the mean values: 
$$
\langle V_t \rangle = (\pi/2)^{1/2} s; \quad 
\langle V_t^2 \rangle = 2s^2; \quad 
\langle V_t^3 \rangle = 3(\pi/2)^{1/2} s^3. 
\eqno(A1) 
$$

In the main text, we assumed that the velocities in $\theta$ and 
$\varphi$ directions are normally distributed with standard 
deviation $s$, so, according to above, the tangential velocity 
$V_t$ density is $\exp{(-v_t^2/2s^2)} v_t/s^2$. This allows to 
write out all the necessary moments of phase density in Eq.~(4), 
which are included in the Jeans Eqs.~(5): 
$$ 
\begin{array}{lll}
	u = nv, \quad p = nv^2, \quad q = 2ns^2, \quad 
	s_{01} = (\pi/2)^{1/2}\,ns, \vspace{2mm} & \\ 
	s_{11} = (\pi/2)^{1/2}\,nvs, \quad s_{12} = 2nvs^2, \quad 
	s_{21} = (\pi/2)^{1/2}\,nv^2s, \vspace{2mm} & \\ 
	s_{30} = nv^3, \quad s_{03} = 3(\pi/2)^{1/2}\,ns^3. 
\end{array} 
\eqno(A2) 
$$
Substituting Eqs.~(A2) into Eqs.~(5) of the main text transforms 
this system into the following: 
$$ 
\left \{    
\begin{array}{lllll} 
	\dfrac{\partial n}{\partial t} + \dfrac{1}{r^2} 
	\dfrac{\partial}{\partial r} (r^2 nv) = 0, \vspace{2mm} & \\ 
	\dfrac{\partial (nv)}{\partial t} + \dfrac{1}{r^2} 
	\dfrac{\partial}{\partial r} (r^2 nv^2) - \dfrac{2ns^2}{r} 
	= -n\, \dfrac{\partial \Phi}{\partial r}, \vspace{2mm} & \\ 
	\dfrac{\partial (ns)}{\partial t} + 
	\dfrac{1}{r^2} \dfrac{\partial}{\partial r} (r^2 nvs) + 
	\dfrac{nvs}{r} = 0, \vspace{2mm} &  \\ 
	\dfrac{\partial (nv^2)}{\partial t} + \dfrac{1}{r^2} 
	\dfrac{\partial}{\partial r} (r^2 nv^3) - \dfrac{4 nvs^2}{r} 
	= -2nv\, \dfrac{\partial \Phi}{\partial r}, \vspace{2mm} & \\ 
	\dfrac{\partial (ns^2)}{\partial t} + 
	\dfrac{1}{r^2} \dfrac{\partial}{\partial r} (r^2 nvs^2) + 
	\dfrac{2 nvs^2}{r} = 0.   \vspace{2mm} \\ 
\end{array}    
\right. 
\eqno(A3) 
$$
The first line of this system is the continuity equation in the 
form of Eq.~(9). Eq.~(A$3_4$), taking into account Eq.~(A$3_2$), 
turns into Eq.~(10). Finally, Eq.~(A$3_5$), taking into account 
Eq.~(A$3_3$), becomes Eq.~(11) of the main text.

\end{document}